\documentclass[showpacs,aps,prb,twocolumn,superscriptaddress]{revtex4}
\usepackage{graphicx} 
\usepackage{dcolumn}
\usepackage{bm}
\usepackage{amssymb,amsmath}
\usepackage{epstopdf}

\begin{document}
\title{Spin Relaxation Times of Single-Wall Carbon Nanotubes}

\normalsize
\author{W.~D.~Rice}
\affiliation{Department of Electrical and Computer Engineering, Rice University, Houston, Texas 77005, USA}
\affiliation{Department of Physics and Astronomy, Rice University, Houston, Texas 77005, USA}
\author{R.~T.~Weber}
\affiliation{Bruker BioSpin Corporation, Billerica, Massachusetts 01821, USA}
\author{P.~Nikolaev}
\affiliation{Air Force Research Lab, Wright-Patterson Air Force Base, Ohio 45433, USA}
\author{S.~Arepalli}
\affiliation{National Institute of Aerospace, 100 Exploration Way, Hampton, Virginia 23666, USA}
\author{V.~Burka}
\author{A.-L.~Tsai}
\affiliation{University of Texas Medical School, Houston, Texas 77005, USA}
\author{J.~Kono}
\email[]{kono@rice.edu}
\thanks{corresponding author.}
\affiliation{Department of Electrical and Computer Engineering, Rice University, Houston, Texas 77005, USA}
\affiliation{Department of Physics and Astronomy, Rice University, Houston, Texas 77005, USA}
\date{\today}

\begin{abstract}
We have measured temperature ($T$)- and power-dependent electron spin resonance in bulk single-wall carbon nanotubes to determine both the spin-lattice and spin-spin relaxation times, $T_1$ and $T_2$.   We observe that $T_1^{-1}$ increases linearly with $T$ from 4 to 100~K, whereas $T_2^{-1}$ {\em decreases} by over a factor of two when $T$ is increased from 3 to 300~K.  We interpret the $T_1^{-1} \propto T$ trend as spin-lattice relaxation via interaction with conduction electrons (Korringa law) and the decreasing $T$ dependence of $T_2^{-1}$ as motional narrowing.  By analyzing the latter, we find the spin hopping frequency to be 285~GHz.  Last, we show that the Dysonian lineshape asymmetry follows a three-dimensional variable-range hopping behavior from 3 to 20~K; from this scaling relation, we extract a localization length of the hopping spins to be $\sim$100~nm.
\end{abstract}


\maketitle



Understanding spin dynamics is key to a broad range of modern problems in condensed matter physics~\cite{SiPRL1997,BalentsPRL2000,KiselevPRB2000,RabelloEuroPhysLett2002,DeMartinoPRL2002,DoraPRL2008} and applied sciences.\cite{DiVincenzoScience1995,WolfScience2001}  Spin transport is a sensitive probe of many-body correlations as well as an indispensable process in spintronic devices.  Confined spins, particularly those in one dimension (1D), are predicted to show strong correlations~\cite{BalentsPRL2000,RabelloEuroPhysLett2002,DeMartinoPRL2002,DoraPRL2008,Giamarchi2004} and long coherence times.\cite{KiselevPRB2000}  Single-wall carbon nanotubes (SWCNTs) are ideal materials for studying 1D spin physics due to their long mean free paths and relatively weak spin-orbit coupling.\cite{AndoJPSJ2000}  Exotic spin properties in metallic SWCNTs at low temperatures and high magnetic fields have been predicted, including the appearance of a peak splitting in the spin energy density spectrum, which can be used to probe spin-charge separation in Luttinger-liquid theory.\cite{RabelloEuroPhysLett2002,DeMartinoPRL2002,DoraPRL2008}

One method for studying spin dynamics is electron spin resonance (ESR), which can provide information on spin-orbit coupling, phase relaxation time, spin susceptibility, and spin diffusion. Many ESR studies of SWCNTs have been performed over the past decade.\cite{DeMartinoPRL2002,DoraPRL2008,KosakaCPL1995,PetitPRB1997,ClayePRB2000,ShenPRB2003,SalvetatPRB2005, NafradiPSS2006, MussoDiaRelMater2006, LikodimosPRB2007, CorziliusPSSB2008, KombarakkaranCPL2008, AngladaPSSB2010, RiceACSNano2012}  Unfortunately, substantial conflicts have emerged in the literature, such as the temperature ($T$) dependence of the spin susceptibility\cite{PetitPRB1997,SalvetatPRB2005, LikodimosPRB2007, RiceACSNano2012} and whether the ESR is caused by SWCNT defects~\cite{KosakaCPL1995, SalvetatPRB2005, MussoDiaRelMater2006, AngladaPSSB2010, RiceACSNano2012} or is intrinsic to nanotubes~\cite{PetitPRB1997, ClayePRB2000, NafradiPSS2006, LikodimosPRB2007, KombarakkaranCPL2008, CorziliusPSSB2008}.  Because of these divergent empirical observations of nanotube ESR, there is only scant experimental data on electron spin-lattice relaxation times in SWCNTs, which limits our understanding of nanotube spin dynamics.

\begin{figure}
\includegraphics [scale = 0.71] {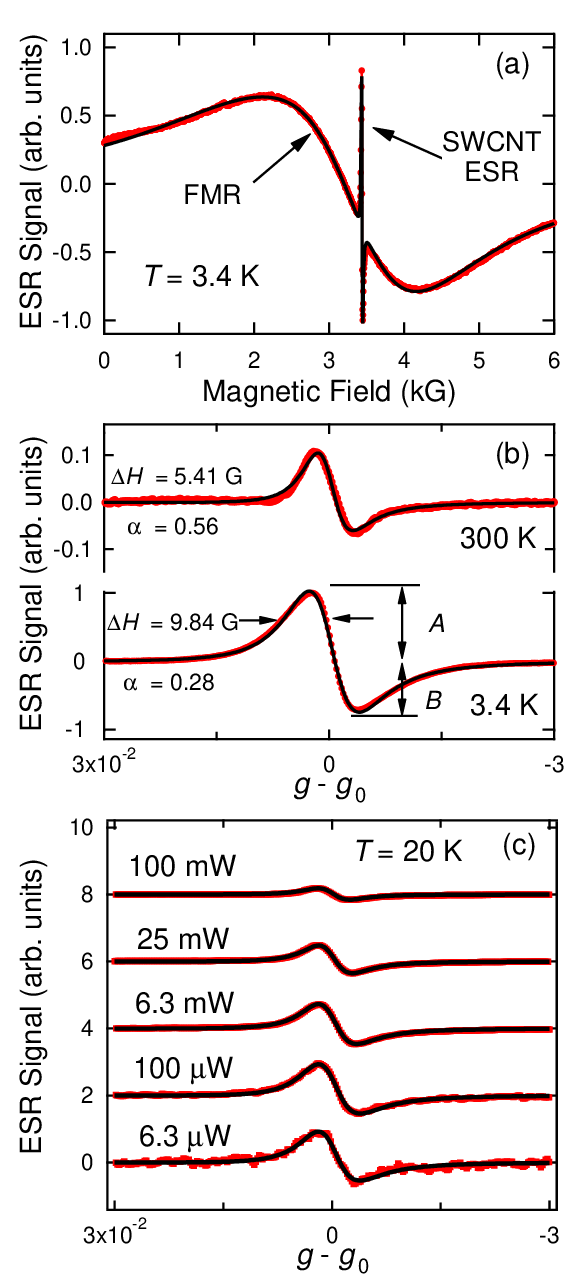}
\caption{(color online) (a)~A raw ESR spectrum taken at 3.4~K across a 6~kG applied field range.  A prominent ESR peak can be seen on top of a large ferromagnetic resonance (FMR) background.  A Dysonian and two Lorentzian line shapes can be used to fully fit the data (black line).  (b)~Background-subtracted ESR scans with Dysonian lineshape fits for the highest (300~K) and lowest temperatures (3.4~K). The Dysonian amplitude and linewidth is clearly much smaller at 300~K than at 3.4~K. (c)~Comparison of ESR traces at 20~K for different microwave powers, showing clear absorption saturation at high powers.  The background was subtracted as in (b), and the spectra are intentionally offset vertically.}
\label{Figure1}
\end{figure}

Here, we present a detailed study of the $T$ dependence of both the spin-lattice ($T_1$) and spin-spin ($T_2$) relaxation times of paramagnetic electron spins in SWCNTs. From the $T$ dependence of $T_1$, we find that the spin-lattice relaxation rate, $T_1^{-1}$, is proportional to $T$.  This trend is consistent with the notion that the probed spins relax through interaction with conduction electrons that are present in metallic SWCNTs in the sample.  Additionally, we find that the dephasing rate, $T_2^{-1}$, becomes smaller as $T$ is increased, which is a hallmark of the phenomenon of motional narrowing.\cite{BloembergenPR1948,KuboJPhysSocJapan1954}  This spin mobility accounts for the Dysonian lineshape\cite{DysonPR1955} seen throughout the full $T$ range examined. The Dysonian lineshape asymmetry parameter, $\alpha$, which is proportional to the conductivity of the probed spins, is shown to follow the 3D variable-range hopping (VRH) trend at low $T$.  


Our sample consisted of acid-purified laser oven SWCNTs in powder form, which we prepared using a comprehensive nanotube compaction and annealing procedure.\cite{RiceACSNano2012} After thermal annealing, the 630~$\mu$g (0.24~g/cm$^3$) SWCNT sample was submerged in mineral oil under helium gas in a sealed quartz tube. To precisely know the value of the perturbing ac magnetic field amplitude, $H_1$, at a given microwave power, we calibrated the dual-mode cavity (Bruker ER4116DM resonator) with a $\alpha$-$\gamma$-bisdiphenylene-$\beta$-phenylallyl (BDPA) complex mixed 1:1 with benzene.  $H_1$ is related to the microwave power, $P$, at a given cavity $Q$ by $H_1 = \alpha_{\rm C}\sqrt{(Q/Q_0)P}$, where $\alpha_{\rm C}$ is the cavity conversion factor and $Q_0$ is the loaded-cavity quality factor during calibration.\cite{Poole1983} To obtain $\alpha_{\rm C}$, we measured the $T_1$ of the BDPA sample using inversion recovery\cite{RabensteinJMagRes1979} to be $T_1$ = 132~ns.  We then performed ESR power saturation spectroscopy on the BDPA calibration sample at $Q_0$ = 5100 and observed that the absorption versus microwave power saturated at 65~mW.  Using this data in conjunction with the $T_1$ and $T_2$ (= 112~ns) values for our BDPA sample, we established $\alpha_{\rm C}$ = 1.83~G/$\sqrt{\rm W}$.

ESR spectra were taken as a function of $T$ from 3.4 to 300~K in the X-band (9.6~GHz) region.  ESR data below 125~K were taken on a Bruker EMX spectrometer using the TE$_{102}$ mode in a dual-mode cavity (Bruker ER4116DM); for $T\geq$~125~K, we used a single mode resonator (Bruker ER4119HS). For the lower $T$ regime, detailed ESR scans were performed with a $P$ of 200~$\mu$W ($H_1\sim1.62\times10^{-2}$~G), while for $T\geq$~125~K, $P$ of 1~mW ($H_1\sim7.31\times10^{-2}$~G) was used.  At certain $T$ values below 125~K, we varied $P$ from 6.3~$\mu$W to 200~mW in steps of 3~dB at an observed $Q$ of 2000~to examine how the relative spin susceptibility changed with $H_1$, as detailed by Portis.\cite{PortisPR1953}

As seen in Fig.~\ref{Figure1}(a), a broad ferromagnetic resonance (FMR) dominates the spectrum, which we attribute to nickel and cobalt catalyst particles remaining in the sample.\cite{LikodimosPRB2007}  A careful study of the lineshape was performed by closely scanning the applied dc magnetic field, $H_0$, around the ESR peak.  Both the linewidth and peak-to-peak amplitude become larger as $T$ is decreased, while the line-center position ($g$-factor, or simply, $g_0$) shows little $T$ dependence.  The ESR line is asymmetric, as seen in Fig.~\ref{Figure1}(b), having what is often referred to as a Dysonian lineshape,\cite{DysonPR1955} indicating movement of the electrons in and out of the $H_1$ perturbing magnetic field.  Figure~\ref{Figure1}(c) shows five traces at 20~K at different microwave powers, spanning more than four orders of magnitude.  As $P$ is increased, the relative ESR signal begins to decrease, as evidenced by the reduction of the signal, normalized for $P$, as $P$ is increased from 6.3~$\mu$W to 100~mW.

To gain further quantitative understanding, we numerically fit each ESR spectrum. The FMR background was fit by a combination of two large-linewidth ($\sim$1000~G) Lorentzians.  The ESR feature was fit using the weak form of the Dysonian lineshape~\cite{SitaramPRB2005, LikodimosPRB2007}
\begin{equation}
\frac{d\chi}{dH_0} = A\chi_{\rm g}\left(\frac{\cos\varphi}{\Delta H_0^2}\right)\frac{-2y + (1-y^2)\tan\varphi}{(1+y^2)^2},
\label{weak Dysonian}
\end{equation}
where $A$ is a coefficient accounting for experimental factors, $\chi_{\rm g}$ is the mass spin susceptibility, $y = \frac{H_0 - H_{\rm r}}{\Delta H_0}$, $H_{\rm r}$ is the resonance field, $\Delta H_0$ is the half-width and is equal to $\frac{1}{\gamma T_2}$, with $\gamma = \frac{g\mu_{\rm B}}{\hbar}$, and $\mu_{\rm B}$ the Bohr magneton.  The weak form of Dysonian can be used here because the conductivity and diffusion of the electrons in the SWCNT powder are both low as compared to a traditional metal.  Nevertheless, unlike traditional magnetic resonance where the signal entirely depends on the imaginary part of the ac spin susceptibility, $\chi''$, the Dysonian lineshape is also influenced by the real component, $\chi'$. Taking the ac susceptibility, $\chi$, to be
\begin{equation}
\chi = \chi''\textrm{cos}\varphi + \chi'\textrm{sin}\varphi,
\label{ac susceptibility}
\end{equation}
we define $\alpha \equiv \tan\varphi$, which is a dimensionless measure of the relative contribution of the real part (= 0 for traditional, fixed spin ESR).  This parameter is also a measure of the asymmetry of the lineshape, and $A/B \approx 1 + \alpha$ when $\alpha \ll 1$, where  $A$ and $B$ are defined in Fig.~\ref{Figure1}(b); additionally, it can be related to the electrical conductivity of the probed spins, $\sigma_{\rm spin}$, as $\alpha \propto \sigma_{\rm spin}$.\cite{SitaramPRB2005}

\begin{figure}
\includegraphics [scale = 0.7] {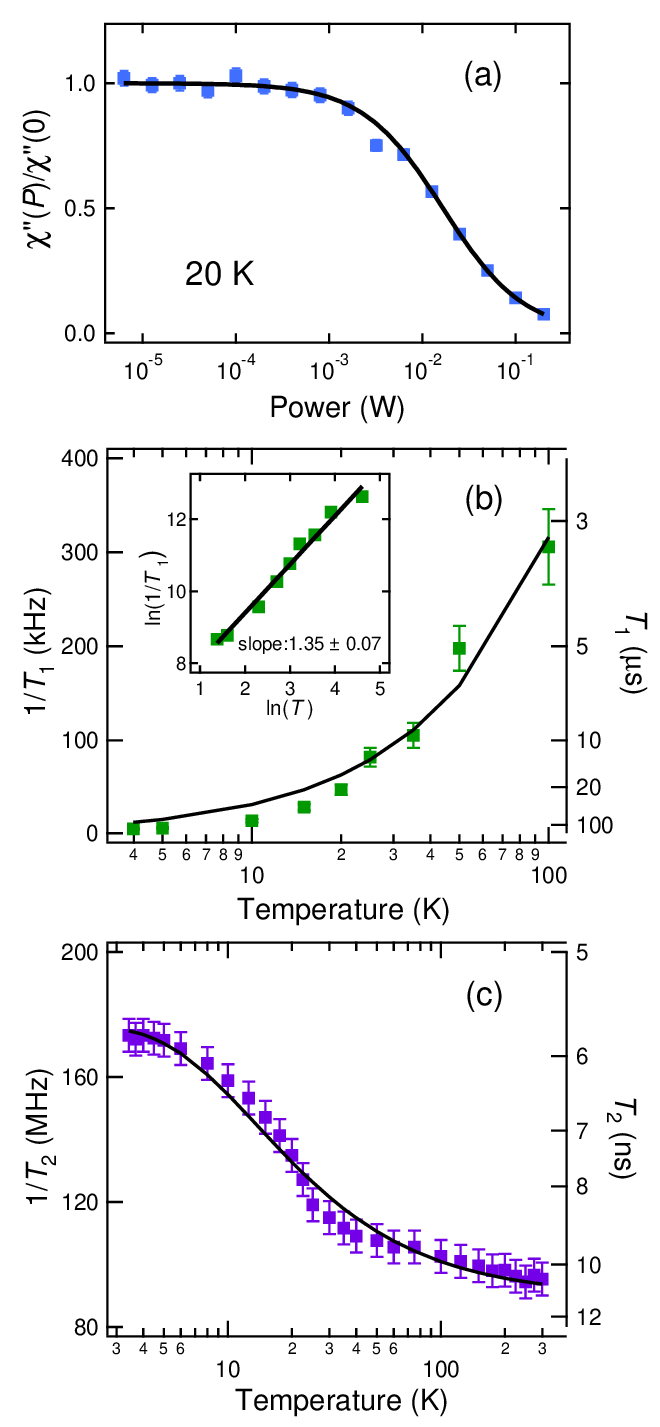} 
\caption{(color online) (a)~Normalized spin susceptibility versus the microwave power, $P$, at 20~K.  The black line shows the fit of Eq.~(\ref{saturation_formula}) to the data. (b)~The $T$ dependence of the spin-lattice relaxation rate, $1/T_1$.  The fit of $1/T_1 = CT$ is shown by the black line. Inset:  A plot of $\ln\left(1/T_1\right)$ versus $\ln\left(T\right)$ shows that data follows a linear relation over the entire measured $T$ range.  (c)~The $T$-dependent spin-spin relaxation rate, $1/T_2$, and the fit of Eq.~(\ref{linewidth eqn}) to the data (black line).}
\label{Figure2}
\end{figure}

From the numerical fitting, we extracted $T_2$, $\alpha$, $g_0$, and $A\chi_{\rm g}$ for each curve as a function of $P$ and $T$.  Since we are in the homogeneous broadening regime, as indicated by the Lorentzian-like lineshape fitting for all curves, we can use the two-level model of $\chi''$:
\begin{equation}
\chi''(P) = \chi_{\rm g}\frac{\omega_{\rm r} T_2}{1 + \left(\omega_0 - \omega_{\rm r}\right)^2T_2^2 + \gamma^2H_1^2T_1T_2},
\label{imaginary_spin_suscept_formula}
\end{equation}
where $H_1 = \alpha_{\rm C}\sqrt{(Q/Q_0)P}$, $\hbar\omega_0 = g\mu_{\rm B}H_0$ and $\omega_{\rm r}$ is the center of the resonance.  At small values of $P$, we can ignore the last term in the denominator, since it will contribute negligibly to the lineshape.  However, as $P$ becomes larger, this saturation term becomes increasingly important, leading to a decrease in $\chi''$.  By taking the ratio of $\chi''\left(P\right)$ to $\chi''\left(P\rightarrow0\right)$ the effect of this saturation term can be clearly delineated:~\cite{PortisPR1953, TothPRB2008}
\begin{equation}
{\chi''\left(P\right) \over \chi''\left(0\right)} = \frac{1}{1 + \gamma^2H_1^2T_1T_2}.
\label{saturation_formula}
\end{equation}
To reduce the experimental errors for the weak signals when $P$ is on the order of 10$^{-5}$~W, we averaged the values of $A\chi_{\rm g}$ (after normalizing for $P$), $T_2$, and $\gamma$ for spectra taken when $P$ was in the linear regime.  The ratio of $\chi''\left(P\right)/\chi''\left(0\right)$ is equivalent to the ratio of $A\chi_{\rm g}\left(P\right)/A\chi_{\rm g}\left(0\right)$, as long as the ESR is not inhomogeneous broadened with increasing $P$.  

A typical plot of $\chi''\left(P\right)/\chi''\left(0\right)$ versus $P$ at 20~K is given in Fig.~\ref{Figure2}(a).  For other $T$'s, Eq.~(\ref{saturation_formula}) also fits well, although minor sample heating effects at the highest powers are seen when $T<10$~K.  From the power saturation fitting, along with the knowledge of $T_2$ and $\gamma$, we can extract $T_1$ for each $T$.   $T_1$ is found to monotonically increase as $T$ decreases [Fig.~\ref{Figure2}(b)]:  when $T$ is lowered from 100 to 4~K, $T_1$ rises from 3.3 to 172~$\mu$s, in agreement with written claims by Clewett \textit{et al.}\cite{ClewettJPhysChemC2007} and the measurements done below 30~K by Musso and co-workers.\cite{MussoDiaRelMater2006}

To better understand the spin-lattice relaxation mechanism, we plotted $\ln\left(1/T_1\right)$ against $\ln\left(T\right)$ [inset of Fig.~\ref{Figure2}(b)]  and observe a nearly linear scaling: $T_1^{-1} \propto T^{1.35\pm0.07}$.  This $T$-dependence closely matches both the Korringa law and direct one-phonon relaxation mechanisms, which go as $T_1^{-1} = CT$, where $C$ is a proportionality constant. As Fig.~\ref{Figure2}(b) shows, the one-variable linear fit follows the general trend of the data well; from this fit, we extract the value of $C$ to be (2.8$\pm$0.4)$\times$10$^3$~sec$^{-1}$K$^{-1}$. 

This $T$-linear behavior of $T_1^{-1}$ is consistent with spin-lattice relaxation via interaction with either conduction electrons (Korringa law) or phonons.\cite{Kuzmany1998}  Since the $g$-factor difference from the free electron value ($\Delta g = g - 2.0023$) suggests small spin-orbit coupling in our system,\cite{RiceACSNano2012} direct spin-phonon interactions are minimal.  In addition, we used a non-enriched SWCNT system, where metallic nanotubes are present.  Thus, we believe that a Korringa law spin-lattice relaxation process is the most likely explanation of the $T_1^{-1} \propto T$ trend.  In this scenario, the probed spins are exchange-coupled to delocalized conduction electrons within $k_{\rm B} T$ of the Fermi level.  Similar conclusions about the spin relaxation were reached in C$_{59}$N-C$_{60}$ heterodimers in an ensemble of non-enriched SWCNTs.\cite{SimonPRL2006, SimonPSSB2007}  However, Musso \textit{et al.}~also saw a linear relationship between $T_1^{-1}$ and $T$ in non-enriched SWCNTs over a limited range (4 to 30~K) but interpreted it in terms of direct phonon relaxation.\cite{MussoDiaRelMater2006}


We also obtained $T_2^{-1}$ from the fitting of the ESR spectra in the linear regime of $H_1$. Unlike previous studies of ESR in nanotubes, $T_2^{-1}$ changes substantially with $T$.  As Fig.~\ref{Figure2}(c) shows, as $T$ is increased from 3.4~K, $T_2^{-1}$ rapidly {\em decreases} until $\sim$25~K, whereupon the dephasing rate begins to decrease more slowly up to 300~K.  This decrease of the ESR linewidth with increasing $T$ is consistent with the phenomenon of motional narrowing,\cite{BloembergenPR1948, KuboJPhysSocJapan1954}~which occurs because the dephasing time of the spins can change as their translational energy is altered.  At high $T$, the spins move rapidly, allowing for less time around dephasing centers, thus reducing the interaction between the probed spins and the dephasing centers.  This decreased interaction gives a longer spin dephasing time ($T_2$), which in turn narrows the lineshape; conversely, at low $T$, the spins are moving more slowly, which broadens the line.

To understand the observation of motional narrowing more quantitatively, we start with a generalized narrowing model~\cite{AndersonRMP1953, AndersonJPSJ1954}
\begin{equation}
T_2^{-1} \simeq \frac{\gamma\Delta H_p^2}{\Delta H_e},
\label{Anderson's relation}
\end{equation}
where $\Delta H_p$ is the amplitude of the perturbations and $\Delta H_e$ describes the rate of spin motion.  Equation~(\ref{Anderson's relation}) was originally derived to describe exchange narrowing or motional narrowing from spin diffusion.  However, spin diffusion can be described in terms of phonon-activated hopping with a probability, $p_{\rm hop}$, that is proportional to $\textrm{exp}\left(-2R/\xi - \Delta E/k_{\rm B}T\right)$.\cite{MillerPR1960,Mott1993,Kamimura1989}  Here, $R$ is the hopping distance, $\xi$ is the localization length, $\Delta E$ is the average spacing between energy levels, and $k_{\rm B}$ is the Boltzmann constant.  Combining this hopping conduction with Eq.~(\ref{Anderson's relation}) and adding an offset, $\left(T_2^0\right)^{-1}$, gives~\cite{WilsonPR1964, MorigakiJPSJ1965}
\begin{equation}
T_2^{-1} = \left(T_2^0\right)^{-1} + \gamma\frac{A}{\Delta E\times\left[1 + \coth\left(\frac{\Delta E}{2k_{\rm B}T}\right)\right]}
\label{linewidth eqn}
\end{equation}
where $T_2^0$ is the high-$T$ (``metallic") asymptotic limit of the spin dephasing time and $A$ is independent of $T$.  As Fig.~\ref{Figure2}(c) shows, Eq.~(\ref{linewidth eqn}) fits very well to the observed linewidth.  We find $A$ to be 11.6$\pm0.8$~meV-G and a $T_2^0$ spin dephasing time of 11.1~ns.  The activation energy, $\Delta E$, is 1.18$\pm$0.09~meV (13.7~K or 285~GHz).  From the value of $\Delta E$, we estimate how much time (on average) each spin spends at each hopping location, $\tau = \frac{\hbar}{\Delta E} = 558$~fs.\cite{KittelISSP1996}  If we phenomenologically take $T_2 = n\tau$, where $n$ is the number of jumps before phase coherence is lost, then we can estimate $n$ to be on the order of 10$^4$ hops, where we have taken $T_2$ to be $\sim$10~ns.

\begin{figure}
\includegraphics [scale = 0.7] {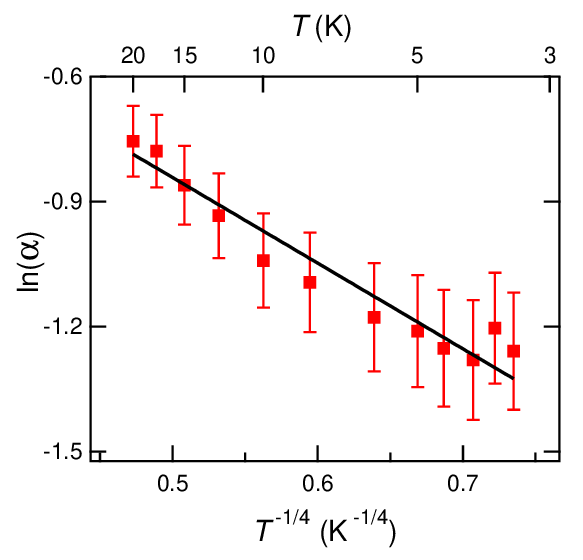}
\caption{(Color online) The natural log of the asymmetry Dysonian parameter, $\alpha$, plotted against the fourth root of inverse temperature.  The trend of ln($\alpha$) follows a 3D VRH behavior, as given by the best fit line in black.}
\label{Figure3}
\end{figure}

To gain deeper insight into the spin hopping mechanism, we examined the asymmetry Dysonian lineshape parameter, $\alpha$, which is proportional to the conductance of the probed spins.  In particular, we were interested to see if $\alpha$ followed a VRH behavior at low $T$, which is mathematically given as~\cite{Mott1993,Kamimura1989}
\begin{equation}
\alpha = \alpha_0\times\textrm{exp}\left[-\left(\frac{T_0}{T}\right)^{\frac{1}{1+d}}\right],
\label{VRH}
\end{equation}
where $T_0$ is the characteristic temperature and $d$ is the dimensionality of the system. As Fig.~\ref{Figure3} shows, ln($\alpha$) follows a linear trend with $T^{-\frac{1}{4}}$, indicating that the spins follow a 3D VRH from 3.4 to 20~K.  From our fit, $T_0$ is 17.9$\pm$5.5~K and $\alpha_0$ is 1.20.  The asymptotic limit of the $\alpha$ parameter, $\alpha_0$, approaches 1 as $T\rightarrow 0$, since the asymmetry of the ESR signal is caused by thermally-activated hopping:  as the phonon density decreases, so does the line shape asymmetry.  The localization length, $\xi$, of the electronic wavefunction can be found from $T_0$~\cite{Kamimura1989}
\begin{equation}
\xi = \left[\frac{18.1}{k_{\rm B}T_0D(E_{\rm F})}\right]^{1/3},
\label{localization length eqn}
\end{equation}
where $D(E_{\rm F})$ is the density of states around the Fermi energy, $E_{\rm F}$.  We can estimate $D(E_{\rm F})$ by treating the defect density of states as having an energy separation that can be roughly estimated by $\Delta E$.  Thus, $D(E_{\rm F}) \approx \frac{N(E_{\rm F})}{\Delta E} \sim 10^{19}$~states/cm$^3$-eV, where we are utilizing the spin density extracted from the Curie constant, $N(E_{\rm F}) =$ 1.14 $\times$ 10$^{16}$~spins/cm$^3$~obtained in our earlier work.\cite{RiceACSNano2012}  From Eq.~(\ref{localization length eqn}), we estimate $\xi$ to be $\sim$100~nm, similar to previous measurements of defect-induced localization lengths in SWCNTs.\cite{GomezNatMat2005}  The spacing of defects, $R_{\rm d}$, can be estimated by $\left(\frac{4\pi}{3}N\right)^{-1/3}$, or $\sim$28~nm.  A $d$ = 3 VRH behavior is expected in this wavefunction-overlap regime, since $R_{\rm d} < \xi$.\cite{Mott1993}  Exchange effects may also be important, but a thorough defect concentration dependence is needed to investigate this avenue more fully. 

It is important to note that given the difficulty in extracting $\alpha$, the $T^{-1/4}$ trend that we are observing can be considered robust.  We also performed traditional conductance measurements on a similarly prepared sample, and although the conductance clearly showed 3D VRH behavior, we believe that our ESR and four-point probe conductivity measurements are probing different species, since the hopping parameters do not agree and the high-$T$ trends are different.  

Although it is clear that the ESR signal arises from nanotubes, its microscopic origin is not certain.  Previously, we advocated that $n$-type defects are essential for the SWCNT ESR, a conclusion based on the observation that ESR signal strongly depends on the presence of molecular oxygen (a $p$-type acceptor), which we attribute to a compensation mechanism.\cite{RiceACSNano2012}  This hypothesis is consistent with the data we present here.  Localized spins that couple via the exchange interaction to conduction electrons would show a $T_1^{-1} \propto T$ scaling behavior.  ESR-active defect states would explain the localized, phonon-assisted hopping of the spins and the motional narrowing of $T_2$; this type of line narrowing was observed in doped semiconductors.  Furthermore, if we take the spin susceptibility value, $\chi_{\rm g} =1.11\pm0.04\times10^{-7}$~emu-K/g and calculate the number of spins per unit cell, assuming a idealized 1~$\mu$m long, (10,10) nanotube (similar to our average diameter), we find that there are 1.4$\times10^{-4}$ ESR-active spins per unit cell, a substantial deviation from the $\approx$1 spin/unit cell expected for an intrinsic SWCNT response.  If this ESR-active defect hypothesis is correct, the wide variety of prior SWCNT ESR results may be due to the different defect concentrations, which would change the $T$ dependences of $T_1$, $T_2$, $\alpha$, and $\chi_{\rm g}$.  

In summary, we have performed temperature- and power-dependent ESR on an ensemble of SWCNTs.  We find $T_1^{-1} \propto T$ from 4 to 100 K, which we interpret as Korringa spin-lattice relaxation.  Furthermore, we observe that $T_2$ undergoes motional narrowing as $T$ is increased from 3.4 to 300~K.  The Dysonian asymmetry parameter, $\alpha$, follows a $T^{-1/4}$ trend at $T \leq 20$~K, which strongly suggests a 3D VRH spin transport at low $T$.  From the extracted parameters, we estimate the spin localization length to be $\sim$100~nm.  These results provide significant new insights into spin relaxation dynamics in SWCNTs.

\begin{acknowledgments}
This work was supported by the DOE/BES (Grant No.~DEFG02-06ER46308), the NSF (Grant Nos.~OISE-0530220 and OISE-0968405), the Robert A.~Welch Foundation (Grant No.~C-1509), the Air Force Research Laboratories (FA8650-05-D-5807), the W.~M.~Keck Program in Quantum Materials at Rice University, the Korean Ministry of Education, Science and Technology under the World Class University Program (R31-2008-10029), and the NIH National Heart, Lung, and Blood Institute (Grant No.~HL095820).  We thank Q.~Si, A.~Imambekov, and R.~Hauge for useful discussions.
\end{acknowledgments}


\end{document}